\documentclass[12pt]{article}
\usepackage{amsmath}
\usepackage{graphicx}
\usepackage{bm}
\usepackage{fancyhdr}
\usepackage{amssymb}
\usepackage{setspace}
\usepackage{epsfig}
\usepackage{overpic}
\usepackage{caption,subcaption}

\allowdisplaybreaks[4]

\begin{document}



\def\a{\alpha}
\def\b{\beta}
\def\d{\delta}
\def\e{\epsilon}
\def\g{\gamma}
\def\h{\mathfrak{h}}
\def\k{\kappa}
\def\l{\lambda}
\def\o{\omega}
\def\p{\wp}
\def\r{\rho}
\def\t{\tau}
\def\s{\sigma}
\def\z{\zeta}
\def\x{\xi}
\def\V={{{\bf\rm{V}}}}
 \def\A{{\cal{A}}}
 \def\B{{\cal{B}}}
 \def\C{{\cal{C}}}
 \def\D{{\cal{D}}}
\def\K{{\cal{K}}}
\def\O{\Omega}
\def\R{\bar{R}}
\def\T{{\cal{T}}}
\def\L{\Lambda}
\def\f{E_{\tau,\eta}(sl_2)}
\def\E{E_{\tau,\eta}(sl_n)}
\def\Zb{\mathbb{Z}}
\def\Cb{\mathbb{C}}

\def\R{\overline{R}}

\def\beq{\begin{equation}}
\def\eeq{\end{equation}}
\def\bea{\begin{eqnarray}}
\def\eea{\end{eqnarray}}
\def\ba{\begin{array}}
\def\ea{\end{array}}
\def\no{\nonumber}
\def\le{\langle}
\def\re{\rangle}
\def\lt{\left}
\def\rt{\right}

\newtheorem{Theorem}{Theorem}
\newtheorem{Definition}{Definition}
\newtheorem{Proposition}{Proposition}
\newtheorem{Lemma}{Lemma}
\newtheorem{Corollary}{Corollary}
\newcommand{\proof}[1]{{\bf Proof. }
        #1\begin{flushright}$\Box$\end{flushright}}

\baselineskip=20pt

\newfont{\elevenmib}{cmmib10 scaled\magstep1}
\newcommand{\preprint}{
   \begin{flushleft}
   \end{flushleft}\vspace{-1.3cm}
   \begin{flushright}\normalsize
   \end{flushright}}
\newcommand{\Title}[1]{{\baselineskip=26pt
   \begin{center} \Large \bf #1 \\ \ \\ \end{center}}}
\newcommand{\Author}{\begin{center}
   \large \bf
Yi Qiao${}^{a,b}$, Zhirong Xin${}^{a}$, Xiaotian Xu${}^{b}$, Kun Hao${}^{a,c}$, Tao Yang${}^{a,c,d}$, Junpeng Cao${}^{b,e,f}\footnote{Corresponding author: junpengcao@iphy.ac.cn}$ and Wen-Li Yang${}^{a,c,d}\footnote{Corresponding author: wlyang@nwu.edu.cn}$

 \end{center}}
\newcommand{\Address}{\begin{center}

     ${}^a$Institute of Modern Physics, Northwest University,
     Xian 710127, China\\
     ${}^b$Beijing National Laboratory for Condensed Matter
           Physics, Institute of Physics, Chinese Academy of Sciences, Beijing
           100190, China\\
     ${}^c$Shaanxi Key Laboratory for Theoretical Physics Frontiers,  Xian 710127, China\\
     ${}^d$School of Physics, Northwest University, Xian 710127, China\\
     ${}^e$School of Physical Sciences, University of Chinese Academy of Sciences, Beijing, China\\
     ${}^f$Songshan Lake Materials Laboratory, Dongguan, Guangdong 523808, China

   \end{center}}

\preprint \thispagestyle{empty}
\bigskip\bigskip\bigskip

\Title{Correlation functions of the XXZ spin chain with the twisted boundary condition} \Author

\Address \vspace{0.1cm}

\vspace{1truecm}

\begin{abstract}

The scalar products, form factors and correlation functions of the XXZ spin chain with twisted (or antiperiodic) boundary condition
are obtained based on the inhomogeneous $T-Q$ relation and the Bethe states constructed via the off-diagonal Bethe
Ansatz. It is shown that the scalar product of two off-shell Bethe states, the form factors and the two-point correlation functions can be expressed as the summation of certain determinants. The corresponding homogeneous limits are studied. The results are also checked by the numerical calculations.

\vspace{1truecm}

\end{abstract}

\newpage


\hbadness=10000

\tolerance=10000

\hfuzz=150pt

\vfuzz=150pt
\section{Introduction}
\setcounter{equation}{0}

The quantum lattice models with integrable boundary conditions have draw a wide attention during these years \cite{Kor93, 3Baxter 2, Tak99}.
The typical quantization conditions of the energy spectrum include the periodic, twisted (or anti-periodic) \cite{Yun95,Bat95,Nie09,Nic13} and open boundary conditions \cite{Che84,Skl88,Fan96,Cao03,Nep04}.
For the integrable models with periodic or diagonal boundary reflection, the Bethe Ansatz methods have been applied successfully \cite{10L.A. Takhtadzhan,Skl78}.
However, for the models with twisted or off-diagonal boundary reflection, the $U(1)$ symmetry of the system is broken and an obvious reference state is missing, which
prevent us from applying the conventional Bethe Ansatz methods to solve the system exactly.
Recently, the off-diagonal Bethe Ansatz (ODBA) is developed to study the exact solutions of quantum integrable models,
especially for those with nontrivial integrable boundary reflections \cite{Wan15}.

The Bethe Ansatz is a powerful method to calculate the partition functions \cite{Ize87} and the correlation functions \cite{Kor93} for one-dimensional integrable systems.
In particular, for models related to the $gl_2$ symmetry, Slavnov \cite{Sla89} and the Gaudin-Korepin \cite{Gau81,Kor82} formulas lead
to that the scalar product between an eigenstate and an arbitrary state, and that the norm of the eigenstates have a determinant representation.
In \cite{Mai00,Goh00}, it is shown the local operator can be expressed by the elements of monodromy matrix, which give a direct path to study the form factors and correlation functions \cite{Kit99}. The thermodynamical limit of the determinant is considered for XXZ model with a diagonal open boundary condition \cite{Kit09}.
Recently, The determinant representations of the scalar products and correlation functions are extensively studied for models related to various boundary conditions especially those with non-diagonal boundary terms \cite{Nic13,Yan11,Yan11-1,Kun12,Nep03,Kit07,Fal14,Bel15}.
Among them, the seperation of variables(SOV) method \cite{Nic13,Skl85,Fra08} gives an attractive approach to obtain the determinant representations of the scalar product between an eigenstate and an arbitrary state with the open boundary conditions.

The XXZ spin chain with twisted boundary condition is a very interesting integrable model \cite{Yun95,Bat95,Nie09,Nic13}.
By using the ODBA method, the spectrum and the inhomogeneous $T-Q$ relation of the system are obtained \cite{Cao13}.
The distribution of Bethe roots at the ground state and the low-energy excitation are studied in the ferromagnetic region \cite{Qia18}.
Based on the inhomogeneous $T-Q$ relations, the Bethe-type eigenstates, which have well defined homogeneous limit,
are retrieved \cite{Zha15}.

In this paper, we study the scalar products, the form factors and the two points correlation functions of the XXZ spin chain with twisted boundary condition.
The corresponding determinant representations are obtained. We also provide the explicit forms of the homogeneous limit and check our results by the numerical calculations.

The paper is organized as follows. Section 2 serves as an introduction of the model and its exact solution. In section 3, we discuss the SoV basis.
In section 4, the Bethe states are reviewed and the scalar product of two off-shell Bethe states are calculated.
In section 5, we present a compact form for the local operators in terms of the elements of the monodromy matrix.
In section 6, form factors are calculated and expressed by certain determinants.
In section 7, we calculate the two points correlation functions.
In section 8, the homogeneous limit of above results are studied.
Numerical results are given in section 9 and the concluding remarks are given in section 10.

\section{XXZ spin chain with twisted boundary conditions}
\setcounter{equation}{0}

The XXZ spin chain with twisted boundary condition is characterized by the Hamiltonian
\bea\label{Ham}
H=-\sum_{j=1}^N\big[ \sigma_j^x \sigma_{j+1}^x + \sigma_j^y \sigma_{j+1}^y +\cosh\eta \sigma_j^z \sigma_{j+1}^z \big],
\eea
where $N$ is the number of sites, $\eta$ is the crossing parameter and the boundary condition is the twisted one, namely,
\bea
\sigma_{N+1}^\alpha=\sigma_1^x\sigma_1^\alpha\sigma_1^x,\,\,{\rm for}\,\, \alpha=x, y, z.\label{Anti-periodic}
\eea
It is remarked that the twisted boundary condition breaks the $U(1)$-symmetry of the system.

The integrability of the model (\ref{Ham}) is associated with the six-vertex $R$-matrix
\begin{eqnarray}\label{R-matrix}
  R_{0,j}(u)= \frac{1}{2} \left[ \frac{\sinh(u+\eta)}{\sinh \eta} (1+\sigma^z_j \sigma^z_0) +\frac{\sinh u}{\sinh \eta} (1- \sigma^z_j \sigma^z_0)   \right]+ \frac{1}{2} (\sigma^x_j \sigma^x_0 +\sigma^y_j \sigma^y_0) ,
\end{eqnarray}
where $u$ is the spectral parameter.
The $R$-matrix satisfies the quantum Yang-Baxter equation (QYBE)
\bea
R_{1,2}(u_1-u_2)R_{1,3}(u_1-u_3)R_{2,3}(u_2-u_3)
=R_{2,3}(u_2-u_3)R_{1,3}(u_1-u_3)R_{1,2}(u_1-u_2).\label{QYB}
\eea
From the $R$-matrix, we can define the monodromy matrix as
\begin{equation}\label{monodromy-matrix}
  T_0(u)=\sigma_0^x R_{0,N}(u-\theta_N) \cdots R_{0,1}(u-\theta_1)=\left(
                                                                     \begin{array}{cc}
                                                                       C(u) & D(u)\\
                                                                       A(u) & B(u)\\
                                                                     \end{array}
                                                                   \right),
\end{equation}
where $\{\theta_j, j=1,\cdots, N\}$ are the inhomogeneous parameters.
Let us introduce further the left quasi-vacuum state $\langle 0|$ and the right quasi-vacuum state $ |0\rangle$
\bea
\langle 0|=\left(
  \begin{array}{cc}
    1, & 0 \\
  \end{array}
\right)_{[1]}\otimes\cdots\otimes\left(
  \begin{array}{ccc}
    1, & 0 \\
  \end{array}
\right)_{[N]},\quad\quad|0\rangle=\left(\begin{array}{c}
    1 \\
    0 \\
  \end{array}\right)_{[1]} \otimes\cdots\otimes\left(\begin{array}{c}
    1 \\
    0 \\
  \end{array}\right)_{[N]} .\label{left-vacuum}
\eea
The elements of monodromy matrix (\ref{monodromy-matrix}) acting on the quasi-vacuum states give rise to
\bea
&&\langle 0|\,A(u)=a(u)\,\langle 0|, \quad \langle 0|\,B(u)=0, \quad \langle 0|\,C(u)\neq 0,\quad D(u)=d(u)\,\langle 0|,\no \\ [4pt]
&&A(u)\,|0\rangle=a(u)\,|0\rangle, \quad B(u)\,|0\rangle\neq 0,\quad C(u)\,|0\rangle=0, \quad D(u)\,|0\rangle=d(u),\label{right-action-1}
\eea
where
\bea\label{ad_func}
a(u)=\prod_{k=1}^{N}\frac{\sinh(u-\theta_k+\eta)}{\sinh \eta}, \quad d(u)=\prod_{k=1}^{N}\frac{\sinh(u-\theta_k)}{\sinh \eta}.
\eea
The operator $B(u)$ [or $C(u)$] acts on the quasi-vaccum states $|0\rangle$ [or $\langle 0 |$] as a creation operator.
The $R$-matrix and the monodromy matrix $T$ satisfy the RTT relation
\begin{equation}
  R_{0,\bar{0} }(u-v) T_0(u)T_{\bar{0}}(v)=T_{\bar{0}}(v) T_0(u) R_{0,\bar{0} }(u-v).\label{hai10}
\end{equation}
From the RTT relation (\ref{hai10}), we obtain the commutative relations among the elements of the monodromy matrix as
\begin{eqnarray}
&&[  B(u),  B(v)]=[  C(u),  C(v)]=0, \label{RLL-1} \no \\[4pt]
&&   A(u)  B(v)=\frac{\sinh(u-v-\eta)}{\sinh(u-v)}  B(v)
  A(u)+\frac{\sinh\eta}{\sinh(u-v)}  B(u)  A(v), \label{RLL-2} \no \\[4pt]
&&   D(u)  B(v)=\frac{\sinh(u-v+\eta)}{\sinh(u-v)}  B(v)  D(u)
-\frac{\sinh\eta}{\sinh(u-v)}  B(u)  D(v), \label{RLL-3} \no \\[4pt]
&&   C(u)  A(v)=\frac{\sinh(u-v+\eta)}{\sinh(u-v)}  A(v)
  C(u)-\frac{\sinh\eta}{\sinh(u-v)}  A(u)  C(v), \label{RLL-4} \no \\[4pt]
&&   C(u)  D(v)=\frac{\sinh(u-v-\eta)}{\sinh(u-v)}  D(v)
  C(u)+\frac{\sinh\eta}{\sinh(u-v)}  D(u)  C(v), \label{RLL-5} \no \\[4pt]
&& [  C(u),  B(v)]=\frac{\sinh\eta}{\sinh(u-v)}[  D(u)  A(v)-  D(v)  A(u)]. \label{RLL-6}
\end{eqnarray}
The transfer matrix of the system is defined as
\begin{equation}\label{transfer_matrix}
t(u)=tr_0 \{ T_0(u) \}=B(u)+C(u).
\end{equation}
Again, from the RTT relation (\ref{hai10}), one can prove that the transfer matrices with different spectral parameters commute with each other,
\begin{equation}
  [t(u),t(v)]=0.
\end{equation}
Then the transfer matrix $t(u)$ serves as the generating functional of all the conserved quantities, which ensures the
integrability of the system. The Hamiltonian (\ref{Ham}) is chosen as the first order derivative of the logarithm of the transfer matrix, namely,
\begin{equation}
  H=-2 \sinh \eta  \frac{\partial \ln t(u)}{\partial u}\big|_{u=0, \{ \theta_j=0 \}} + N \cosh \eta. \label{hai11}
\end{equation}

The transfer matrix (\ref{transfer_matrix}) and the Hamiltonian (\ref{Ham}) can be exactly solved by using the ODBA method \cite{Wan15, Cao13}.
Let $|\Phi\rangle$ be the common eigenstates of the transfer matrix (\ref{transfer_matrix}) and the Hamiltonian (\ref{Ham}) with the eigenvalues $\Lambda(u)$ and $E$, respectively,
\bea
t(u)|\Phi\rangle =\Lambda(u)|\Phi\rangle,\quad H |\Phi\rangle =E |\Phi\rangle.\nonumber
\eea
The eigenvalue $\Lambda(u)$ can be parameterized into following inhomogeneous $T-Q$ relation
\bea\label{Lambda}
\Lambda(u)Q(u)=a(u)e^u Q(u-\eta)-e^{-u-\eta}d(u)Q(u+\eta)-c(u)a(u)d(u),
\eea
where $Q(u)$ is a trigonometric polynomial with the form
\bea
Q(u)=\prod_{j=1}^{N} \frac{\sinh(u-\lambda_j)}{\sinh\eta},
\eea
the $\{\lambda_j\}$ are the Bethe roots and the coefficient $c(u)$ is given by
\bea
c(u)=e^{u-N\eta+\sum_{l=1}^{N}(\theta_l-\lambda_l)}-e^{-u-\eta-\sum_{l=1}^{N}(\theta_l-\lambda_l)}.
\eea
The $N$ Bethe roots $\{ \lambda_j \}$ satisfy the Bethe Ansatz equations (BAEs)
\bea\label{BAEs}
e^{\lambda_j}a(\lambda_j)Q(\lambda_j-\eta)-e^{-\lambda_j-\eta}d(\lambda_j)Q(\lambda_j+\eta)
-c(\lambda_j)a(\lambda_j)d(\lambda_j)=0, \quad j=1,\cdots,N.
\eea
The eigen-energy of the Hamiltonian (\ref{Ham}) is then expressed in terms of the Bethe roots as
\bea\label{energy}
E=2\,\sinh\eta\sum_{j=1}^N \big[ \coth(\lambda_j+\eta)-\coth(\lambda_j) \big]
-N\cosh\eta-2\sinh\eta.
\eea

\section{SoV basis}
\setcounter{equation}{0}

A convenient SoV basis of the model (\ref{Ham}) is \cite{Nic13}
\bea
&& |h_{1},\cdots,h_{N} \rangle = \prod_{j=1}^{N}[ B(\theta_{j}) ]^{h_j} | 0 \rangle,\label{basis_2} \\
&& \langle h_{1},\cdots,h_{N}| = \langle 0 | \prod_{j=1}^{N}[ C(\theta_{j}) ]^{h_j}, \label{basis_12}
\eea
where $h_j=0$ or $1$.
Using the commutation relations (\ref{RLL-6}), one can derive following orthogonal relations between the right states (\ref{basis_2}) and the left states (\ref{basis_12})
\bea\label{det_f}
f(h) &=& \langle h_{1},\cdots,h_{N}|h_{1},\cdots,h_{N} \rangle  \no \\
&=& \prod_{l=1}^{N}[-a(\theta_l)d(\theta_l-\eta)e^{-\eta(N-1)} ]^{h_l}\frac{| V(\theta_1, \cdots, \theta_N) |}{| V(\theta_1-\eta h_1 ,\cdots, \theta_N-\eta h_N) |},
\eea
where $|V|$ means the determinant of matrix $V$ which is given by
\bea
    V(x_1,\cdots,x_N)=\left( \begin{array}{cccc}
                                1& e^{2x_1}& \cdots& e^{2x_1(N-1)}\\
                                1& e^{2x_2}& \cdots& e^{2x_2(N-1)}\\
                                \vdots & \vdots& \  &\vdots\\
                                1& e^{2x_N}& \cdots& e^{2x_N(N-1)}\\
                             \end{array}
                       \right). \label{hai}
\eea
The above determinant is a kind of Vandermonde determinant, which possesses the following properties
\bea\label{Vand_1}
&&|V(x_1,\cdots,x_N)|=\prod_{1\leq j<i\leq N}(e^{2x_i}-e^{2x_j}), \no \\
&&|V(x_1,\cdots,x_N,y)|=\prod_{1\leq j<i\leq N}(e^{2x_i}-e^{2x_j})\prod_{k=1}^{N}(e^{2y}-e^{2x_k}).
\eea
The right states (\ref{basis_2}) form an orthogonal right basis of the Hilbert space, while
the left states (\ref{basis_12}) form an orthogonal left basis of the Hilbert space. Thus we have
\bea
{\rm id} = \sum_{\{h\}}|h_{1},\cdots,h_{N} \rangle \langle h_{1},\cdots,h_{N}| f^{-1}(h), \label{identity}
\eea
where ${\rm id}$ is the unitary matrix, $\sum_{\{h\}}$ means sum over all possible values of $\{h_1\cdots h_N\}$, and the total number of the terms is $2^N$.

We note that the SoV basis (\ref{basis_2}) is the eigenstate of the operator $D(u)$,
\bea
D(u)|h_{1},\cdots,h_{N} \rangle
=\prod_{j=1}^{N} \frac{\sinh(u-\theta_j+\eta h_j)}{\sinh\eta}|h_{1},\cdots,h_{N} \rangle.
\eea
Meanwhile, we also know the action of operator $C(u)$ acting on SoV basis (\ref{basis_2})
\bea\label{C_expansion}
&&\hspace{-1.2truecm} C(u)|h_{1},\cdots,h_{N} \rangle
=\sum_{i=1}^{N} \prod_{j\neq i}^{N}
\frac{\sinh(u-\theta_j+\eta h_j)}{ \sinh\eta }
\frac{\sinh(\theta_i-\theta_j-\eta h_j)}{\sinh(\theta_i-\theta_j)}a(\theta_i) \no \\
&&\hspace{2.5truecm} \times |h_{1},\cdots,h_i-1,\cdots,h_{N} \rangle,
\eea
which will be used later.

\section{Bethe states and scalar products}
\setcounter{equation}{0}
\subsection{Bethe states}

The Bethe states of the system read
\bea\label{bethe_state}
|\lambda_1,\cdots,\lambda_N\rangle &=& \prod_{j=1}^{N}D(\lambda_j) | \Omega; \{\theta_j\} \rangle, \\
\langle \lambda_1,\cdots,\lambda_N| &=& \langle \Omega; \{\theta_j\} | \prod_{j=1}^{N}D(\lambda_j),
\eea
where $|\Omega; \{\theta_j\} \rangle $ is the reference state which can be determined by the inner products
\bea \label{ref_state}
&& \langle h_{1},\cdots,h_{N}| \Omega; \{\theta_j\} \rangle = \prod_{l=1}^{N} [ a(\theta_{l})d(\theta_{l}) ]^{h_l}, \\
&& \langle \Omega; \{\theta_j\} | h_{1},\cdots,h_{N} \rangle = \prod_{l=1}^{N} [ a(\theta_{l})d(\theta_{l}) ]^{h_l}.
\eea
We express the Bethe states (\ref{bethe_state}) by the SoV basis (\ref{basis_2}) as
\bea
|\lambda_1,\cdots,\lambda_N\rangle
&=& \sum_{h}|h_{1},\cdots,h_{N} \rangle \langle h_{1},\cdots,h_{N}| f^{-1}(h)
|\lambda_1,\cdots,\lambda_N\rangle  \no \\
&=& \sum_{h} f^{-1}(h) \prod_{j=1}^{N} \bar d(\{\lambda_k\},\theta_j,h_j) a^{h_j}(\theta_j)e^{h_j\theta_j}
\ |h_{1},\cdots,h_{N} \rangle ,
\eea
where
\bea\label{dbar}
\bar d(\{\lambda_k\},u,h)=\prod_{k=1}^{N}\frac{\sinh(\lambda_k-u+\eta h)}{\sinh \eta}.
\eea

If the Bethe roots $\{\lambda_j, j = 1, \cdots , N \}$ satisfy the BAEs (\ref{BAEs}),
then the Bethe states (\ref{bethe_state}) are the common eigenstates of the Hamiltonian (\ref{Ham}) and the transfer matrix (\ref{transfer_matrix}).
The eigenstates, denoted as $|\Phi\{\lambda_k\}\rangle$, can also be expressed by the SoV basis (\ref{basis_2}) as
\bea
|\Phi\{\lambda_k\}\rangle
&=& \sum_{h}f^{-1}(h)|h_{1},\cdots,h_{N} \rangle \langle h_{1},\cdots,h_{N}|\Phi\{\lambda_k\}\rangle
\no \\
&=& \sum_{h} f^{-1}(h) \prod_{j=1}^{N}\Lambda^{h_j}(\{\lambda_k\},\theta_j)
\ |h_{1},\cdots,h_{N} \rangle.
\eea
In the derivation, we have used following identities
\bea
&&\langle h_{1},\cdots,h_{N}|\Phi\{\lambda_k\}\rangle=\prod_{j=1}^{N}d(u_j)\Lambda^{h_j}(\{\lambda_k\},\theta_j), \no \\
&&\langle \Phi\{u_k\} |h_{1},\cdots,h_{N} \rangle =\prod_{j=1}^{N}d(u_j)\Lambda^{h_j}(\{u_k\},\theta_j), \no \\
&& \bigg[\frac{\sinh(\theta_i-\theta_j+\eta)}{\sinh(\theta_i-\theta_j)}\bigg]^{h}=\frac{\sinh(\theta_i-\theta_j+\eta h)}{\sinh(\theta_i-\theta_j)},
\no \\[4pt]
&&\frac{\sinh(\theta_i-\theta_j-\eta h)\sinh(\theta_i-\theta_j-\eta+\eta h)}{\sinh(\theta_i-\theta_j)}=\sinh(\theta_i-\theta_j-\eta),
\eea
where $h = 0$ or $1$.

\subsection{Scalar product of two off-shell Bethe states}

Now, we consider the scalar product of two off-shell Bethe states. After tedious calculation, we obtain
\bea \label{det_scalar product0}
&&\hspace{-1.8cm}  S_N\equiv\langle u_1,\cdots,u_N|\lambda_1,\cdots,\lambda_N\rangle
= \langle \Omega; \{\theta_j\} |   \prod_{k=1}^{N}D(u_k) D(\lambda_k) | \Omega; \{\theta_j\} \rangle \no \\
&\stackrel{(\ref{identity})}{=}&\sum_h  f^{-1}(h) \prod_{j=1}^{N} \bigg\{   d(u_j)d(\lambda_j) [a^2(\theta_j)e^{2\theta_j}]^{h_j} \prod_{k=1}^{N} \frac{\sinh(u_j-\theta_k+\eta h_k)}{\sinh(u_j-\theta_k)}  \no \\ &&
\quad \times \frac{\sinh(\lambda_j-\theta_k+\eta h_k)}{\sinh(\lambda_j-\theta_k)}  \bigg\}.
\eea
We also find that above result can be expressed by the ratio of two determinants
\bea\label{det_scalar product}
S_N=\frac{|P|}{|V(\theta_1,\cdots,\theta_N)|},
\eea
where $V(\theta_1,\cdots,\theta_N)$ is given by Eq.(\ref{hai}), $P$ is the $N\times  N$ matrix with the elements
\bea\label{det_scalar product_ele}
P_{ij}=\sum_{h=0}^1 \tau( \{u_l\},\{\lambda_l\},h,\theta_i)e^{2(\theta_i-\eta h)(j-1)},
\eea
and the function $\tau(\{u_l\},\{\lambda_l\},h,u)$ is
\bea\label{tau}
\tau(\{u_l\},\{\lambda_l\},h,u) = \bar d(\{u_k\},u,h)
\bar d(\{\lambda_k\},u,h)
\bigg[ \frac{-a(u)}{d(u-\eta)} \bigg]^{h}
e^{2u h+\eta h(N-1)}.
\eea

\subsection{Scalar product of one off-shell Bethe state and one on-shell Bethe state}

Next, we consider the scalar product of an eigenstate and a off-shell Bethe state.
Suppose the parameters $\{u_k\}$ satisfy the BAEs (\ref{BAEs}), and we obtain
\bea \label{det_norm}
\hspace{-2cm}&& \langle \Phi\{u_k\} |\lambda_1,\cdots,\lambda_N\rangle \no \\
\hspace{-0.8cm}&\stackrel{(\ref{identity})}{=}& \sum_h  f^{-1}(h) \prod_{j=1}^{N} \bigg\{   d(u_j)d(\lambda_j) [a(\theta_j)e^{\theta_j}\Lambda(\{u_k\},\!\theta_j)]^{h_j} \prod_{k=1}^{N} \frac{\sinh(\lambda_j-\theta_k+\eta h_k)}{\sinh(\lambda_j-\theta_k)}  \bigg\}.
\eea
We find that above equation can be expressed as
\bea\label{det_norm_left}
\langle \Phi\{u_k\} |\lambda_1,\cdots,\lambda_N\rangle=\frac{|P^{NL}|}{|V(\theta_1,\cdots,\theta_N)|},
\eea
where $V(\theta_1,\cdots,\theta_N)$ is given by Eq.(\ref{hai}) and $P^{NL}$ is a $N\times  N$ matrix with the elements
\bea\label{det_norm_ele}
P^{NL}_{ij}=\sum_{h=0}^1 e^{\eta h(N-1)+\theta_i h +2(\theta_i-\eta h)(j-1)}d(u_i)  \left[\frac{-\Lambda(\{u_k\},\theta_i)}{d(\theta_i-\eta)} \right]^{h}
\prod_{k=1}^{N}\frac{\sinh(\lambda_k-\theta_i+\eta h)}{\sinh\eta}.
\eea

If the right Bethe state is a eigenstate and the left Bethe state is the off-shell one, the scalar product reads
\bea\label{det_norm_right}
\langle u_1,\cdots,u_N|\Phi\{\lambda_k\}\rangle=\frac{|P^{NR}|}{|V(\theta_1,\cdots,\theta_N)|},
\eea
where $P^{NR}$ is a $N\times  N$ matrix with the elements $P^{NR}_{ij}=P^{NL}_{ij}|_{\lambda_k \leftrightarrow u_k}$.

\section{Inverse problem}
\setcounter{equation}{0}

The inverse problem is to reconstruct the local operators by the elements of monodromy matrix.
The general method to solve the inverse problem of the quantum integrable models was proposed in \cite{Mai00,Goh00}.
The inverse problem of the closed XXZ spin chain was considered in \cite{Kit99} and the antiperiodic case was considered in \cite{Nic13}.
Here, the main procedures to solve the quantum inverse problem are reviewed and a slightly different form of the results for the antiperiodic case are obtained (see (\ref{inverseD})-(\ref{inverseZ}) below).

At the point of $u=0$, the $R$-matrix (\ref{R-matrix}) degenerates into the permutation operator, $R_{0,j}(0)= P_{0,j}$, which possesses the property
\begin{eqnarray}
R_{i,j}(u)=P_{0,i}R_{0,j}(u)P_{0,i}. \label{hai7}
\end{eqnarray}
Considering the value of transfer matrix $t(u)$ at the point of $u=\theta_j$ and using the property (\ref{hai7}), we obtain
\begin{eqnarray}\label{t-theta1}
&&\hspace{-0.5in}t(\theta_j) = tr_0\{\sigma^x_0 R_{0,N}(\theta_j-\theta_N)\cdots R_{0,j+1}(\theta_j-\theta_{j+1})P_{0,j}
R_{0,j-1}(\theta_j-\theta_{j-1})\cdots R_{0,1}(\theta_j-\theta_1)\} \no \\
&&\hspace{-0.1in}= R_{j,j-1}(\theta_j-\theta_{j-1})\cdots R_{j,1}(\theta_j-\theta_1)\sigma^x_j
R_{j,N}(\theta_j-\theta_N)\cdots R_{j,j+1}(\theta_j-\theta_{j+1}),\no \\
&&\hspace{-0.1in} \equiv (j,j-1)\cdots(j,1)\sigma^x_j(j,N)\cdots(j,j+1),
\end{eqnarray}
where we have used the notation $(i,j)\equiv R_{i,j}(\theta_i-\theta_j)$. The production of $t(\theta_j)$ gives
\begin{eqnarray}\label{times-t-theta1}
\prod^{i}_{j=1}t(\theta_j)
\!&=&\!\prod^{i-1}_{j=1}\prod^{i}_{k>j}\phi_{jk}\cdot\sigma^x_1\cdots\sigma^x_i[(1,N)\cdots(1,i+1)][(2,N)\cdots(2,i+1)]\cdots
\no \\ &&\times [(i,N)\cdots(i,i+1)],
\end{eqnarray}
where
\begin{eqnarray}\label{phi_ij}
\phi_{jk}=\sinh(\eta - \theta_j+\theta_k)\sinh(\eta + \theta_j-\theta_k)/\sinh^2\eta.
\end{eqnarray}
Because the transfer matrices with different spectral parameters commute with each other,
the inverse of quantity (\ref{times-t-theta1}) is
\begin{eqnarray}\label{times-t-theta2}
\!\!\!\!\prod^{i}_{j=1} t^{-1}(\theta_j)
\!\!&=&\!\!\prod^{i-1}_{j=1}\prod^{i}_{k>j}\phi^{-1}_{jk}\cdot\prod^{i}_{m=1}\prod^{N}_{n=i+1}\phi^{-1}_{mn}
\cdot [(i+1,i)\cdots(N,i)]\cdots[(i+1,2)\cdots(N,2)]\no\\
&&\!\! \times [(i+1,1)\cdots(N,1)]\sigma^x_1\cdots\sigma^x_i.
\end{eqnarray}
Let $i=N$ in (\ref{times-t-theta1}), we obtain following identities
\begin{eqnarray}
   &&\!\!\!\!\prod^{N}_{j=1} t(\theta_j)=\prod^{N-1}_{i=1}\prod^{N}_{j>i}\phi_{ij}\cdot \sigma^x_1\cdots\sigma^x_N, \\
   &&\!\!\!\!\prod^{N}_{j=1} t(\theta_j)\cdot\prod^{N}_{k=1} t(\theta_k)=\prod^{N-1}_{i=1}\prod^{N}_{j>i}\phi^2_{ij}\times{\rm id}.\label{t-identity}
 \end{eqnarray}
By using Eqs.(\ref{times-t-theta1}) and (\ref{times-t-theta2}), we also have
\begin{eqnarray}\label{right-side}
&&
\prod^{i-1}_{j=1}t^{-1}(\theta_{j})\, x_i \prod^{i}_{j=1}t(\theta_{j})
=[(i,i-1)\cdots(i,1)]x_i\sigma^x_i[(i,N)\cdots(i,i+1)],
\end{eqnarray}
where $x_i$ is a local operator defined in $i$-th space.
Meanwhile,
\begin{eqnarray}\label{left-side}
tr_0\{x_0T_0(\theta_i)\} &=&
tr_0\{x_0\sigma^x_0 R_{0,N}(\theta_i-\theta_N)\cdots
R_{0,i+1}(\theta_i-\theta_{i+1}) P_{0,i}R_{0,i-1}(\theta_i-\theta_{i-1})\cdots  \no \\ \no &&
\times R_{0,1}(\theta_i-\theta_1) \} \\ \no
&=& R_{i,i-1}(\theta_i-\theta_{i-1})\cdots R_{i,1}(\theta_i-\theta_1)x_i\sigma^x_i R_{1,N}(\theta_1-\theta_N)\cdots R_{i,i+1}(\theta_i-\theta_{i+1}) \\
&=&[(i,i-1)\cdots(i,1)]x_i\sigma^x_i[(i,N)\cdots(i,i+1)].
\end{eqnarray}
Comparing Eqs.(\ref{right-side}) and (\ref{left-side}), we arrive at the identity
\bea\label{inverse 1}
tr_0\{x_0 T_0(\theta_i)\}=\prod^{i-1}_{j=1}t^{-1}(\theta_{j})  x_i \prod^{i}_{j=1}t(\theta_{j}).
\eea
With the help of Eqs.(\ref{t-identity}) and (\ref{inverse 1}), we express the generators of $su(2)$ algebra in terms of elements of the monodromy matrix as
\begin{eqnarray}
&&  \sigma_{i}^-=\prod^{N-1}_{i=1}\prod^{N}_{j>i}\phi^{-2}_{ij} \cdot \prod^{i-1}_{j=1}t(\theta_{j})\cdot D(\theta_i)\cdot \prod^{N}_{j=i+1}t(\theta_{j})
  \cdot \prod^{N}_{j=1}t(\theta_{j}),\label{inverseD}
 \\
&&  \sigma_{i}^+=\prod^{N-1}_{i=1}\prod^{N}_{j>i}\phi^{-2}_{ij} \cdot \prod^{i-1}_{j=1}t(\theta_{j})\cdot A(\theta_i)\cdot \prod^{N}_{j=i+1}t(\theta_{j})
  \cdot \prod^{N}_{j=1}t(\theta_{j}),\label{inverseA}
\\
&&  \sigma_{i}^z=\prod^{N-1}_{i=1}\prod^{N}_{j>i}\phi^{-2}_{ij} \cdot \prod^{i-1}_{j=1}t(\theta_{j})\cdot [2C(\theta_i)-t(\theta_i)]\cdot \prod^{N}_{j=i+1}t(\theta_{j})
  \cdot \prod^{N}_{j=1}t(\theta_{j}),\label{inverseZ}
\end{eqnarray}
which will be used to calculate the form factors and the correlation functions.

\section{Form factors}
\setcounter{equation}{0}

Suppose $\langle\Phi\{u_k\}|$ is the left eigenstate of the model (\ref{Ham}) while $|\Phi\{\lambda_k\}\rangle$ is the right eigenstate.
First, we calculate the form factor $\langle\Phi\{u_k\}|\sigma_i^-|\Phi\{\lambda_k\}\rangle$. Using the formula (\ref{inverseD}), we obtain
\bea
\langle\Phi\{u_k\}|\sigma_i^-|\Phi\{\lambda_k\}\rangle
&=&\prod^{N-1}_{i=1}\prod^{N}_{j>i}\phi^{-2}_{ij} \prod^{i-1}_{j=1}\Lambda(\{u_k\},\theta_{j}) \prod^{N}_{j=i+1}\Lambda(\{\lambda_k\},\theta_{j}) \prod^{N}_{j=1}\Lambda(\{\lambda_k\},\theta_{j})   \no \\
&&\times \langle\Phi\{u_k\}| D(\theta_i) |\Phi\{\lambda_k\}\rangle.
\eea
The factor $\langle\Phi\{u_k\}| D(\theta_i) |\Phi\{\lambda_k\}\rangle$ reads
\bea
&& \hspace{-1cm}
\langle\Phi\{u_k\}| D(\theta_i) |\Phi\{\lambda_k\}\rangle
\stackrel{(\ref{identity})}{=}\sum_{\{h\}} \prod_{j=1}^{N} \bigg\{ d(u_j)d(\lambda_j) \bigg[ \frac{-a(\theta_j)e^{\eta(N-1)+2 \theta_j}}{d(\theta_j-\eta)} \bigg]^{h_j}\frac{\sinh(\theta_i- \theta_j+\eta h_j)}{\sinh \eta} \no\\ 
&&\quad \times \prod_{k=1}^{N} \frac{\sinh(u_j-\theta_k+\eta h_k)}{\sinh(u_j-\theta_k)}
\frac{\sinh(\lambda_j-\theta_k+\eta h_k)}{\sinh(\lambda_j-\theta_k)}  \bigg\}
\frac{| V(\theta_1-\eta h_1 ,\cdots, \theta_N-\eta h_N) |}{| V(\theta_1, \cdots, \theta_N) |},
\eea
which can be written as the determinant expression
\bea\label{det_D}
\langle\Phi\{u_k\}| D(\theta_i) |\Phi\{\lambda_k\}\rangle
=\frac{|F^D|}{|V(\theta_1,\cdots,\theta_N)|},
\eea
where $F^D$ is the $N\times  N$ matrix with the elements
\bea
F^D_{m,n}=\sum_{h=0}^1 \tau(\{u_l\},\{\lambda_l\},h,\theta_m)e^{2(\theta_m-\eta h)(n-1)}\frac{\sinh(\theta_i-\theta_m+\eta h ) }{ \sinh \eta}.
\eea
Therefore, we arrive at
\bea
\langle\Phi\{u_k\}|\sigma_i^-|\Phi\{\lambda_k\}\rangle
&=&\prod^{N-1}_{m=1}\prod^{N}_{n>m}\phi^{-2}_{mn} \prod^{i-1}_{j=1}\Lambda(\{u_k\},\theta_{j}) \prod^{N}_{j=i+1}\Lambda(\{\lambda_k\},\theta_{j}) \prod^{N}_{j=1}\Lambda(\{\lambda_k\},\theta_{j})  \no  \\
&&\times \frac{|F^D|}{|V(\theta_1,\cdots,\theta_N)|}.
\eea

Similarly, the form factor $\langle\Phi\{u_k\}| \sigma_i^+ |\Phi\{\lambda_k\}\rangle$ is related with the scalar product
$\langle\Phi\{u_k\}| A(\theta_i) |\Phi\{\lambda_k\}\rangle$, which can be calculated directly and the results are very complicated and omitted here.
We note that the form factor $\langle\Phi\{u_k\}| \sigma_i^+ |\Phi\{\lambda_k\}\rangle$ can also be calculated by a simple way as follows.
First one can construct another set of SoV basis from the eigenstates of the operator $A(u)$ in (\ref{monodromy-matrix}). Then
using this basis and following the same procedure as above, one can obtain the form factor $\langle\Phi\{u_k\}| \sigma_i^+ |\Phi\{\lambda_k\}\rangle$.

Next, we consider the $\langle\Phi\{u_k\}|\sigma_i^z|\Phi\{\lambda_k\}\rangle$. By using the reconstruction of local operator $\sigma_i^z$ (\ref{inverseZ}), we have
\bea
\langle\Phi\{u_k\}|\sigma_i^z|\Phi\{\lambda_k\}\rangle
&=&\prod^{N-1}_{m=1}\prod^{N}_{n>m}\phi^{-2}_{mn} \prod^{i-1}_{j=1}\Lambda(\{u_k\},\theta_{j}) \prod^{N}_{j=i+1}\Lambda(\{\lambda_k\},\theta_{j}) \prod^{N}_{j=1}\Lambda(\{\lambda_k\},\theta_{j})   \no \\
&&\times \langle\Phi\{u_k\}| \left[2C(\theta_i)- \Lambda( \{ u_k \},\theta_{i} )\right] |\Phi\{\lambda_k\}\rangle. \label{hai3}
\eea
With the help of Eq.(\ref{C_expansion}), we obtain
\bea
&& \hspace{-2cm}
\langle\Phi\{u_k\}| C(\theta_i) |\Phi\{\lambda_k\}\rangle \stackrel{(\ref{identity})}{=} \sum_{j=1}^{N}{\sum_{\{h\}}}^{\prime}
(-1)^{j+N} \xi(\{u_l\},\{\lambda_l\},\theta_i,\theta_j) \prod_{k\neq j}^{N} \tau(\{u_l\},\{\lambda_l\},h_k,\theta_k)\no\\ 
&& \times
\frac{| V_j(\theta_1-\eta h_1 ,\cdots, \theta_N-\eta h_N, \theta_1) |}{| V(\theta_1, \cdots, \theta_N) |}, \label{hai2}
\eea
where ${\sum_{\{h\}}}^{\prime}$ indicates that the parameter $h_j$ is not included in the
summation of $\{h_1\cdots h_N\}$ and $h_j=1$. $ V_j(x_1,\cdots,x_N,y) $ is the $N \times  N $ matrix with the form
\bea
    V_j(x_1,\cdots,x_N,y)=\left( \begin{array}{cccc}
                                1& e^{2x_1}& \cdots& e^{2x_1(N-1)}\\
                                1& e^{2x_2}& \cdots& e^{2x_2(N-1)}\\
                                \vdots & \vdots& \  &\vdots\\
                                1& e^{2x_{j-1}}& \cdots& e^{2x_{j-1}(N-1)}\\
                                1& e^{2x_{j+1}}& \cdots& e^{2x_{j+1}(N-1)}\\
                                \vdots & \vdots& \  &\vdots\\
                                1& e^{2x_N}& \cdots& e^{2x_N(N-1)}\\
                                1& e^{2y}& \cdots& e^{2y(N-1)}\\
                             \end{array}
                       \right),
\eea
and
\bea
&&\hspace{-2cm}
\xi(\{u_l\},\{\lambda_l\},\theta_i,\theta_j)=\bar d(\{u_k\},\theta_j,1) \bar d(\{\lambda_k\},\theta_j,1)
\bigg[ \frac{-a(\theta_j)}{d(\theta_j-\eta)} \bigg] e^{\theta_i + \eta N}  \no \\
&&\quad \times \prod_{k=1}^{N}\bigg[e^{-\theta_i+\theta_j-\eta}
\frac{\sinh(\theta_j-u_k)}{\sinh(\theta_j-u_k-\eta)}
\frac{\sinh(\theta_j-\theta_k-\eta)}{\sinh \eta}\bigg].
\eea
We find that the scalar product (\ref{hai2}) also has a determinant expression
\bea\label{det_C}
\langle\Phi\{u_k\}| C(\theta_i) |\Phi\{\lambda_k\}\rangle
 =\frac{|F^C|}{|V(\theta_1,\cdots,\theta_N)|},
\eea
where $F^C$ is the $ (N+1) \times  (N+1) $ matrix with the elements
\bea\label{det_C_ele}
&&F^C_{m,n}=\sum_{h=0}^1 \tau(\{u_l\},\{\lambda_l\},h,\theta_m)e^{2(\theta_m-\eta h)(n-1)},
\quad 1\leq m\leq N, 1\leq n\leq N, \no  \\ \no
&&F^C_{m,N+1}=\xi(\{u_l\},\{\lambda_l\},\theta_i,\theta_m), \quad 1\leq m\leq N, \\ \no
&&F^C_{N+1,n}=e^{2\theta_i(n-1)}, \quad 1\leq n\leq N, \\
&&F^C_{N+1,N+1}=0.
\eea
Substituting Eqs.(\ref{det_C}) into (\ref{hai3}), we obtain
\bea\label{det_z}
\langle\Phi\{u_k\}| \sigma_{i}^{z} |\Phi\{\lambda_k\}\rangle
&=&\prod^{N-1}_{m=1}\prod^{N}_{n>m}\phi^{-2}_{mn} \prod^{i-1}_{j=1}\Lambda(\{u_k\},\theta_{j}) \prod^{N}_{j=i+1}\Lambda(\{\lambda_k\},\theta_{j}) \prod^{N}_{j=1}\Lambda(\{\lambda_k\},\theta_{j})   \no \\
&&\times  \frac{|F^z|}{|V(\theta_1,\cdots,\theta_N)|},
\eea
where $F^z$ is the $ (N+1) \times  (N+1) $ matrix with the elements
\bea \label{det_z_ele}
&&F^z_{m,n}=\sum_{h=0}^1 \tau(\{u_l\},\{\lambda_l\},h,\theta_m)e^{2(\theta_m-\eta h)(n-1)},
\quad 1\leq m\leq N, \quad 1\leq n\leq N, \no \\ \no
&&F^z_{m,N+1}=2\xi(\{u_l\},\{\lambda_l\},\theta_i,\theta_m), \quad 1\leq m\leq N, \\ \no
&&F^z_{N+1,n}=e^{2\theta_i(n-1)}, \quad 1\leq n\leq N, \\
&&F^z_{N+1,N+1}=-\Lambda(\{\lambda_k\},\theta_{i}).
\eea

\section{Correlation functions}
\setcounter{equation}{0}

Now, we consider the two points correlation function $\langle\Phi\{u_k\}|\sigma_{i-1}^- \sigma_i^-|\Phi\{\lambda_k\}\rangle$. According to Eq.(\ref{inverseD}), we have
\bea
\langle\Phi\{u_k\}|\sigma_{i-1}^-\sigma_i^-|\Phi\{\lambda_k\}\rangle
&=&\prod^{N-1}_{m=1}\prod^{N}_{n>m}\phi^{-2}_{mn} \prod^{i-2}_{j=1}\Lambda(\{u_k\},\theta_{j}) \prod^{N}_{j=i+1}\Lambda(\{\lambda_k\},\theta_{j}) \prod^{N}_{j=1}\Lambda(\{\lambda_k\},\theta_{j})  \no  \\  &&\times \langle\Phi\{u_k\}| D(\theta_{i-1})D(\theta_i) |\Phi\{\lambda_k\}\rangle.
\eea
Direct calculation shows
\bea
&& \hspace{-2cm}
\langle\Phi\{u_k\}| D(\theta_{i-1})D(\theta_i) |\Phi\{\lambda_k\}\rangle\no\\
&\stackrel{(\ref{identity})}{=}&
\sum_{\{h\}} \prod_{j=1}^{N} \{ d(u_j)d(\lambda_j) \bigg[ \frac{-a(\theta_j)e^{\eta(N-1)+2 \theta_j}}{d(\theta_j-\eta)} \bigg]^{h_j}\frac{\sinh(\theta_i- \theta_j+\eta h_j)}{\sinh \eta} \no\\  && \times \frac{\sinh(\theta_{i-1}-\theta_j+\eta h_j)}{\sinh \eta}
\prod_{k=1}^{N} \frac{\sinh(u_j-\theta_k+\eta h_k)}{\sinh(u_j-\theta_k)}
\frac{\sinh(\lambda_j-\theta_k+\eta h_k)}{\sinh(\lambda_j-\theta_k)}  \} \no \\
&&\times \frac{| V(\theta_1-\eta h_1 ,\cdots, \theta_N-\eta h_N) |}{| V(\theta_1, \cdots, \theta_N) |},
\eea
which can be expressed as the determinant form
\bea\label{det_DD}
\langle\Phi\{u_k\}| D(\theta_{i-1})D(\theta_i) |\Phi\{\lambda_k\}\rangle
=\frac{|F^{DD}|}{|V(\theta_1,\cdots,\theta_N)|},
\eea
where $F^{DD}$ is the $N\times  N$ matrix with the elements
\bea
F^{DD}_{m,n}=\sum_{h=0}^1 \tau(\{u_l\},\{\lambda_l\},h,\theta_m)e^{2(\theta_m-\eta h)(n-1)}\frac{\sinh(\theta_{i-1}-\theta_m+\eta h ) }{\sinh \eta}\frac{\sinh(\theta_i-\theta_m+\eta h ) }{ \sinh \eta}.
\eea
Then we arrive at
\bea
\langle\Phi\{u_k\}|\sigma_{i-1}^-\sigma_i^-|\Phi\{\lambda_k\}\rangle
&=&\prod^{N-1}_{m=1}\prod^{N}_{n>m}\phi^{-2}_{mn} \prod^{i-2}_{j=1}\Lambda(\{u_k\},\theta_{j}) \prod^{N}_{j=i+1}\Lambda(\{\lambda_k\},\theta_{j})   \no \\
&&\times \prod^{N}_{j=1}\Lambda(\{\lambda_k\},\theta_{j}) \frac{|F^{DD}|}{|V(\theta_1,\cdots,\theta_N)|}. \label{hai6}
\eea

Next, we consider the two points correlation function $\langle\Phi\{u_k\}|\sigma_{i-1}^z \sigma_i^z |\Phi\{\lambda_k\}\rangle$.
From Eq.(\ref{inverseZ}), we have
\bea
&&
\langle\Phi\{u_k\}|\sigma_{i-1}^z\sigma_i^z|\Phi\{\lambda_k\}\rangle
=\prod^{N-1}_{m=1}\prod^{N}_{n>m}\phi^{-2}_{mn} \prod^{i-2}_{j=1}\Lambda(\{u_k\},\theta_{j}) \prod^{N}_{j=i+1}\Lambda(\{\lambda_k\},\theta_{j}) \prod^{N}_{j=1}\Lambda(\{\lambda_k\},\theta_{j})   \no \\
&&\quad \times \langle\Phi\{u_k\}| \left\{4C(\theta_{i-1})C(\theta_i)-2C(\theta_{i-1})t(\theta_i)
-t(\theta_{i-1})[2C(\theta_i)-t(\theta_i)]\right\}|\Phi\{\lambda_k\}\rangle.\label{cf_zz}
\eea
Direct calculation shows
\bea \label{cf_CC}
&& \hspace{-2cm}
\langle\Phi\{u_k\}| C(\theta_{i-1})C(\theta_i) |\Phi\{\lambda_k\}\rangle  \no \\ \no
&\stackrel{(\ref{identity})}{=}&\! \sum_{j'=1}^{N}\sum_{j\neq j'}^{N} {\sum_{\{h\}}}^{\prime\prime} \frac{(-1)^{j+j'+1+x}}
{\sinh(\theta_j-\theta_{j'}-\eta)}
\gamma_1(\{u_l \},\{\lambda_l \},\theta_i,\theta_{j'})\gamma_2(\{u_l \},\{\lambda_l \},\theta_i,\theta_{j}) \\
&&\! \times \! \prod_{k\neq j,j'}^{N} \tau(\{u_l\},\{\lambda_l\},h_k,\theta_k)\frac{|V_{j',j}(\theta_1-\eta h_1 ,\cdots, \theta_N-\eta h_N, \theta_{i-1},\theta_{i})|}{| V(\theta_1, \cdots, \theta_N) |},
\eea
where $x=0$ if $j'>j$; $x=1$ if $j'< j $, ${\sum_{\{h\}}}^{\prime\prime}$ means that the parameters $h_{j'}$ and $h_j$ are not included in the summation, $h_{j'}=1$ and $h_{j}=1$,
$ V_{j',j}(x_1,\cdots,x_N,y,z) $ is the $N \times  N $ matrix
\bea
    V_{j',j}(x_1,\cdots,x_N,y,z)=\left( \begin{array}{cccc}
                                1& e^{2x_1}& \cdots& e^{2x_1(N-1)}\\
                                1& e^{2x_2}& \cdots& e^{2x_2(N-1)}\\
                                \vdots & \vdots& \  &\vdots\\
                                1& e^{2x_{j'-1}}& \cdots& e^{2x_{j'-1}(N-1)}\\
                                1& e^{2x_{j'+1}}& \cdots& e^{2x_{j'+1}(N-1)}\\
                                \vdots & \vdots& \  &\vdots\\
                                1& e^{2x_{j-1}}& \cdots& e^{2x_{j-1}(N-1)}\\
                                1& e^{2x_{j+1}}& \cdots& e^{2x_{j+1}(N-1)}\\
                                \vdots & \vdots& \  &\vdots\\
                                1& e^{2x_N}& \cdots& e^{2x_N(N-1)}\\
                                1& e^{2y}& \cdots& e^{2y(N-1)}\\
                                1& e^{2z}& \cdots& e^{2z(N-1)}\\
                             \end{array}
                       \right),
\eea
and the functions $\gamma_1$ and $\gamma_2$ are given by
\bea
&&\gamma_1(\{u_l \},\{\lambda_l \},\theta_i,\theta_{j'})= -\frac{\sinh(\theta_{i-1}-\theta_{j'})}{\sinh(\theta_i-\theta_{i-1})}
\xi(\{u_l\},\{\lambda_l\},\theta_i,\theta_{j'}),
\no \\
&&\gamma_2(\{u_l \},\{\lambda_l \},\theta_i,\theta_{j})=
-\sinh(\theta_{i}-\theta_{j}+\eta) \xi(\{u_l\},\{\lambda_l\},\theta_{i-1},\theta_{j}).
\eea
Again, Eq.(\ref{cf_CC}) can be rewritten as the determinant expression
\bea\label{det_cf_CC}
\langle\Phi\{u_k\}| C(\theta_{i-1})C(\theta_i) |\Phi\{\lambda_k\}\rangle
=\sum_{j'=1}^{N} (-1)^{N+j'}\gamma_1(\{u_l \},\{\lambda_l \},\theta_i,\theta_{j'})
\frac{|F_{j'}^{CC}|}{| V(\theta_1, \cdots, \theta_N) |},
\eea
where $F_{j'}^{CC}$ is the reduced matrix of $F^{CC}$ with $j'$ row vanished and $F^{CC}$ is the $(N+2) \times  (N+1) $ matrix with the elements
\bea \label{det_CC}
&&F^{CC}_{m,n}=\sum_{h=0}^1 \tau(\{u_l\},\{\lambda_l\},h,\theta_m)e^{2(\theta_m-\eta h)(n-1)},
\quad 1\leq m\leq N, 1\leq n\leq N, \no \\ \no
&&F^{CC}_{m,N+1}=\gamma_2(\{u_l \},\{\lambda_l \},\theta_i,\theta_{m})/\sinh(\theta_m-\theta_{j'}-\eta), \quad 1\leq m\leq N, \\ \no
&&F^{CC}_{N+1,n}=e^{2\theta_{i-1}(n-1)}, \quad 1\leq n\leq N, \\ \no
&&F^{CC}_{N+2,n}=e^{2\theta_i(n-1)}, \quad 1\leq n\leq N, \\
&&F^{CC}_{n,N+1}=0, \quad n=N,N+1.
\eea
With the help of Eqs.(\ref{det_scalar product}) and (\ref{det_C}), we have
\bea\label{cf_2}
\langle\Phi\{u_k\}|C(\theta_{i-1})t(\theta_i)|\Phi\{\lambda_k\}\rangle =
\Lambda(\{\lambda_k\},\theta_{i})\frac{|F^C(\theta_{i-1})|}{|V(\theta_1,\cdots,\theta_N)|},
\eea
and
\bea\label{cf_3}
\langle\Phi\{u_k\}|t(\theta_{i-1})[2C(\theta_i)-t(\theta_i)]|\Phi\{\lambda_k\}\rangle =
\Lambda(\{u_k\},\theta_{i-1})\frac{|F^z(\theta_{i})|}{|V(\theta_1,\cdots,\theta_N)|}.
\eea
Substituting Eqs.(\ref{det_cf_CC}), (\ref{cf_2}) and (\ref{cf_3}) into (\ref{cf_zz}), we obtain
\bea\label{det_cfzz}
&& \hspace{-2cm}
\langle\Phi\{u_k\}|\sigma_{i-1}^z\sigma_i^z|\Phi\{\lambda_k\}\rangle
=\prod^{N-1}_{m=1}\prod^{N}_{n>m}\phi^{-2}_{mn} \prod^{i-2}_{j=1}\Lambda(\{u_k\},\theta_{j}) \prod^{N}_{j=i+1}\Lambda(\{\lambda_k\},\theta_{j}) \prod^{N}_{j=1}\Lambda(\{\lambda_k\},\theta_{j})  \no  \\ \no &&\times  \big \{ 4\sum_{j'=1}^{N} (-1)^{N+j'}\gamma_1(\{u_l \},\{\lambda_l \},\theta_i,\theta_{j'}) |F_{j'}^{CC}| -2\Lambda(\{\lambda_k\},\theta_{i})|F^C_{i-1}| \\
&&-\Lambda(\{u_k\},\theta_{i-1}) |F^z|  \big \}  | V(\theta_1, \cdots, \theta_N) |^{-1},
\eea
where $F^C_{i-1}$ is  the $ (N+1) \times  (N+1) $ matrix with the elements
\bea
&&(F^C_{i-1})_{m,n}=\sum_{h=0}^1 \tau(\{u_l\},\{\lambda_l\},h,\theta_m)e^{2(\theta_m-\eta h)(n-1)},
\quad 1\leq m\leq N, \quad 1\leq n\leq N, \no  \\ \no
&&(F^C_{i-1})_{m,N+1}=\xi(\{u_l\},\{\lambda_l\},\theta_{i-1},\theta_m), \quad 1\leq m\leq N, \\ \no
&&(F^C_{i-1})_{N+1,n}=e^{2\theta_{i-1}(n-1)}, \quad 1\leq n\leq N, \\
&&(F^C_{i-1})_{N+1,N+1}=0.
\eea

\section{Homogeneous limit}
\setcounter{equation}{0}

In this section, we consider the the homogeneous limit of above results by using the method suggested in \cite{Ize92}.
From Eq.(\ref{Vand_1}), the denominator of scalar product (\ref{det_scalar product}) is
\bea
|V(\theta_1,\cdots,\theta_N)|=\prod_{1\leq j<i\leq N}(e^{2\theta_i}-e^{2\theta_j}).
\eea
We define
\bea
\varphi_j(u,\{\theta_k\})&=&\sum_{h=0}^1 \tau( \{u_l\},\{\lambda_l\},h,u)e^{2(u-\eta h)(j-1)}, \label{hai4}
\eea
where the function $\tau$ is given by Eq.(\ref{tau}).
Then the scalar product (\ref{det_scalar product}) reads
\bea\label{det_scalar product_2}
 S_N =\left| \begin{array}{cccc}
        \varphi_1(\theta_1,\{\theta_k\})&\varphi_2(\theta_1,\{\theta_k\}) & \cdots& \varphi_N(\theta_1,\{\theta_k\})\\
        \varphi_1(\theta_2,\{\theta_k\})&\varphi_2(\theta_2,\{\theta_k\}) & \cdots& \varphi_N(\theta_2,\{\theta_k\})\\
        \vdots & \vdots& \  &\vdots\\
        \varphi_1(\theta_N,\{\theta_k\})&\varphi_2(\theta_N,\{\theta_k\}) & \cdots& \varphi_N(\theta_N,\{\theta_k\})\\
     \end{array}
 \right| \prod_{j<i}(e^{2\theta_i}-e^{2\theta_j})^{-1}.
\eea
Taking the Taylor expansion of function $\varphi_j(u,\{\theta_k\})$ at the point of $\theta_1$, we obtain
\bea
\varphi_j(u,\{\theta_k\})&=&\varphi_j(\theta_1,\{\theta_k\})
+(u-\theta_1)\varphi_j'(u,\{\theta_k\})|_{u=\theta_1}\no \\
&&+\frac{1}{2}(u-\theta_1)^2 \varphi_j^{(2)}(u,\{\theta_k\})|_{u=\theta_1}+\cdots.\label{phi_expan}
\eea
Put $u=\theta_2$ in Eq.(\ref{phi_expan}) and we have
\bea\label{phi_expan_1}
\varphi_j(\theta_2,\{\theta_k\})=\varphi_j(\theta_1,\{\theta_k\})
+(\theta_2-\theta_1)\varphi_j'(u,\{\theta_k\})|_{u=\theta_1}+\cdots.
\eea
Substitute (\ref{phi_expan_1}) into (\ref{det_scalar product_2}) and taking the limit $\theta_2 \rightarrow \theta_1$, we obtain
\bea\label{det_scalar product_hom2}
S_N &=& \left| \begin{array}{cccc}
        \varphi_1(\theta_1,\{\theta_k\})&\varphi_2(\theta_1,\{\theta_k\}) & \cdots& \varphi_N(\theta_1,\{\theta_k\})\\
        \varphi'_1(u,\{\theta_k\})|_{u=\theta_1}&\varphi'_2(u,\{\theta_k\})|_{u=\theta_1} & \cdots& \varphi'_N(u,\{\theta_k\})|_{u=\theta_1}\\
        \vdots & \vdots& \  &\vdots\\
        \varphi_1(\theta_N,\{\theta_k\})&\varphi_2(\theta_N,\{\theta_k\}) & \cdots& \varphi_N(\theta_N,\{\theta_k\})\\
     \end{array}
 \right| \no \\&& \times \left[2 e^{2\theta_1} \!\!\! \prod_{2\leq j<i\leq N} \!\!(e^{2\theta_i}-e^{2\theta_j})\right]^{-1}. \label{det_scalar product_3}
\eea
Let $u=\theta_3$ and Eq.(\ref{phi_expan}) reads
\bea
\varphi_j(\theta_3,\{\theta_k\})&=&\varphi_j(\theta_1,\{\theta_k\})
+(\theta_3-\theta_1)\varphi_j'(u,\{\theta_k\})|_{u=\theta_1} \no \\
&&+\frac{1}{2}(\theta_3-\theta_1)^2\varphi_j^{(2)}(u,\{\theta_k\})|_{u=\theta_1}+\cdots.\label{phi_expan_2}
\eea
Substitute (\ref{phi_expan_2}) into (\ref{det_scalar product_hom2}) and taking the limit $\theta_3 \rightarrow \theta_1$, we obtain
\bea
S_N&=&\left| \begin{array}{cccc}
        \varphi_1(\theta_1,\{\theta_k\})&\varphi_2(\theta_1,\{\theta_k\}) & \cdots& \varphi_N(\theta_1,\{\theta_k\})\\
        \varphi'_1(u,\{\theta_k\})|_{u=\theta_1}&\varphi'_2(u,\{\theta_k\})|_{u=\theta_1} & \cdots& \varphi'_N(u,\{\theta_k\})|_{u=\theta_1}\\
        \varphi^{(2)}_1(u,\{\theta_k\})|_{u=\theta_1}&\varphi^{(2)}_2(u,\{\theta_k\})|_{u=\theta_1} & \cdots& \varphi^{(2)}_N(u,\{\theta_k\})|_{u=\theta_1}\\
        \vdots & \vdots& \  &\vdots\\
        \varphi_1(\theta_N,\{\theta_k\})&\varphi_2(\theta_N,\{\theta_k\}) & \cdots& \varphi_N(\theta_N,\{\theta_k\})\\
     \end{array}
 \right| \no \\&& \times \left[2(2 e^{2\theta_1})^3 \prod_{3\leq j<i\leq N}(e^{2\theta_i}-e^{2\theta_j})\right]^{-1}.
\eea
Repeat above procedures and we arrive at
\bea
 S_N&=&\left| \begin{array}{cccc}
        \varphi_1(\theta_1,\{\theta_k\})&\varphi_2(\theta_1,\{\theta_k\}) & \cdots& \varphi_N(\theta_1,\{\theta_k\})\\
        \varphi'_1(u,\{\theta_k\})|_{u=\theta_1}&\varphi'_2(u,\{\theta_k\})|_{u=\theta_1} & \cdots& \varphi'_N(u,\{\theta_k\})|_{u=\theta_1}\\
        \varphi^{(2)}_1(u,\{\theta_k\})|_{u=\theta_1}&\varphi^{(2)}_2(u,\{\theta_k\})|_{u=\theta_1} & \cdots& \varphi^{(2)}_N(u,\{\theta_k\})|_{u=\theta_1}\\
        \vdots & \vdots& \  &\vdots\\
        \varphi^{(N-1)}_1(u,\{\theta_k\})|_{u=\theta_1}&\varphi^{(N-1)}_2(u,\{\theta_k\})|_{u=\theta_1} & \cdots& \varphi^{(N-1)}_N(u,\{\theta_k\})|_{u=\theta_1}\\
      \end{array}
\right| \no \\&& \times \left[(2 e^{2\theta_1})^{N(N-1)/2}\prod_{k=1}^{N-1} k!\right]^{-1}.
\eea
Finally, let all the inhomogeneous parameters be zero and we obtain the scalar product (\ref{det_scalar product}) in the homogeneous limit
\bea\label{homo_det_scalar product}
S_N = |P^{hom}|  \left[2^{N(N-1)/2}\prod_{k=1}^{N-1} k!\right]^{-1}.
\eea
Here the elements of matrix $P^{hom}$ are
\bea
&&P^{hom}_{m,n}=\frac {\partial^{m-1} \varphi_n(u)}{\partial u^{m-1}}\bigg|_{u=0}, \no \\
&& \varphi_n(u)=\sum_{h=0}^1 \tilde{\tau}( \{u_l\},\{\lambda_l\},h,u)e^{2(u-\eta h)(n-1)}, \label{hai5}
\eea
where
\bea\label{homo_tau}
\tilde{\tau}(\{u_l\},\{\lambda_l\},h,u) = \bar d(\{u_k\},u,h)
\bar d(\{\lambda_k\},u,h)
\bigg[ \frac{-\sinh^{ N}(u+\eta)}{\sinh^{ N}(u-\eta)} \bigg]^{h}
e^{2u h+\eta h(N-1)},
\eea
and $\bar{d}(\{\lambda_k\},u,h)$ is given by Eq.(\ref{dbar}). We note that the scalar product (\ref{homo_det_scalar product})
is valid for both the on-shell and the off-shell Bethe states, which means that the values of
parameters $\{u_k\}$ and $\{\lambda_k\}$ are arbitrary.

Using the same procedure, the homogeneous limit of the form factors and correlation functions can be obtained.
After some algebras, we have
\bea
\langle\Phi\{u_k\}|\sigma_i^-|\Phi\{\lambda_k\}\rangle
= |F^{hom,D}| \Lambda^{i-1}(\{u_k\},0) \Lambda^{2N-i}(\{\lambda_k\},0)  \left[2^{N(N-1)/2}\prod_{k=1}^{N-1} k!\right]^{-1},
\eea
where the elements of matrix $F^{hom,D}$ are
\bea
&&F^{hom,D}_{m,n}=\frac {\partial^{m-1} f^-_n(u)}{\partial u^{m-1}}\bigg|_{u=0}, \no\\
&&f^-_n(u)=\sum_{h=0}^1 \tilde{\tau}(\{u_l\},\{\lambda_l\},h,u)e^{2(u-\eta h)(n-1)}\frac{\sinh(-u+\eta h ) }{ \sinh \eta}.
\eea
The homogeneous limit of the form factor (\ref{det_z}) is
\bea\label{homo_det_z}
\langle\Phi\{u_k\}| \sigma_{i}^{z} |\Phi\{\lambda_k\}\rangle
= |F^{hom,z}| \Lambda^{i-1}(\{u_k\},0) \Lambda^{2N-i}(\{\lambda_k\},0) \left[2^{N(N-1)/2}\prod_{k=1}^{N-1} k!\right]^{-1},
\eea
where $F^{hom,z}$ is the $ (N+1) \times  (N+1) $ matrix with the elements
\bea \label{homo_det_z_ele}
&&F^{hom,z}_{m,n}=\frac {\partial^{m-1} \varphi_n(u)}{\partial u^{m-1}}\bigg|_{u=0},
\quad 1\leq m\leq N, \quad 1\leq n\leq N, \no \\ \no
&&F^{hom,z}_{m,N+1}=\frac {2 \partial^{m-1} \tilde{\xi}(u)}{\partial u^{m-1}}\bigg|_{u=0}, \quad 1\leq m\leq N, \\ \no
&&F^{hom,z}_{N+1,n}=1, \quad 1\leq n\leq N, \\
&&F^{hom,z}_{N+1,N+1}=-\Lambda(\{\lambda_k\},0),
\eea
the $ \varphi_n(u)$ is given by Eq.(\ref{hai5}) and the $\tilde{\xi}(u)$ reads
\bea\label{tilde_xi}
\tilde{\xi}(u) = -\bar d(\{u_k\},u,1) \bar d(\{\lambda_k\},u,1) e^{u N}
\frac{\sinh^{ N}(u+\eta)}{\sinh^{ N}\eta} \prod_{k=1}^{N}
\bigg[ \frac{\sinh(u-u_k)}{\sinh(u-u_k-\eta)} \bigg].
\eea

The homogeneous limit of the correlation function (\ref{hai6}) is
\bea
&&\langle\Phi\{u_k\}|\sigma_{i-1}^-\sigma_i^-|\Phi\{\lambda_k\}\rangle \no \\
&&\qquad =|F^{hom,DD}| \Lambda^{i-2}(\{u_k\},0) \Lambda^{2N-i}(\{\lambda_k\},0) \left[2^{N(N-1)/2}\prod_{k=1}^{N-1} k!\right]^{-1}, \label{hai9}
\eea
where elements of matrix $F^{hom,DD}$ are
\bea
&&F^{hom,DD}_{m,n}=\frac {\partial^{m-1} f^{--}_n(u)}{\partial u^{m-1}}\bigg|_{u=0}, \no \\
&&f^{--}_n(u)=\sum_{h=0}^1 \tilde{\tau}(\{u_l\},\{\lambda_l\},h,u)e^{2(u-\eta h)(n-1)}\frac{\sinh^2(u-\eta h ) }{ \sinh^2 \eta}.
\eea
The homogeneous limit of the correlation function (\ref{det_cfzz}) is quite complicated to calculate. Here we give it at $N=2$ case
\bea
&& \hspace{-2cm}
\langle\Phi\{u_k\}|\sigma_{1}^z\sigma_2^z|\Phi\{\lambda_k\}\rangle
=\Lambda^{2}(\{\lambda_k\},0)  \no  \\  &&\times  \big \{ 4\tilde{\xi}^2(0) -\Lambda(\{\lambda_k\},0)|F^{hom,C}| -\Lambda(\{u_k\},0) |F^{hom,z}|/2  \big \},
\eea
where $F^{hom,C}$ is the $ 3 \times 3$ matrix with the elements
\bea \label{homo_det_C}
&&F^{hom,C}_{m,n}=\frac {\partial^{m-1} \varphi_n(u)}{\partial u^{m-1}}\bigg|_{u=0},
\quad 1\leq m\leq 2, \quad 1\leq n\leq 2, \no \\ \no
&&F^{hom,C}_{m,3}=\frac {\partial^{m-1} \tilde{\xi}(u)}{\partial u^{m-1}}\bigg|_{u=0}, \quad 1\leq m\leq 2, \\ \no
&&F^{hom,C}_{3,n}=1, \quad 1\leq n\leq 2, \\
&&F^{hom,C}_{3,3}=0,
\eea
where the $ \varphi_n(u)$ is given by Eq.(\ref{hai5}) and the $\tilde{\xi}(u)$ is given by (\ref{tilde_xi}).

\section{Numerical results}
\setcounter{equation}{0}

In this section, we check above results by the numerical calculation. For simplicity, we consider the homogeneous case.
All the scalar products, the form factors and the correlation functions can be obtained by two ways. One is from the definition and the
other is from the analytical formula provided in previous sections. Therefore, we can compare these two results and check the valid of the analytical formula.

Put all the inhomogeneous parameters be zero in Eq.(\ref{bethe_state}), and we obtain the Bethe states in the homogeneous limit.
We note that during this process, the singularity does not arise.
From Eq.(\ref{ref_state}), we obtain the homogeneous right reference state $|\Omega\rangle$
\bea
|\Omega\rangle=\sum_{l=0}^{\infty}\frac{\lt(B^{-}\rt)^l}{[l]_q!}|0\rangle
=\sum_{l=0}^{N}\frac{\lt(B^{-}\rt)^l}{[l]_q!}|0\rangle,\label{reference-state-r}
\eea where the $q$-integers $\{[l]_q, l=0,\cdots\}$ and the operator $B^{-}$ are given by
\bea
&&[l]_q=\frac{1-q^{2l}}{1-q^{2}},\quad [0]_q=1, \no \\
&&[l]_q!=[l]_q\,[l-1]_q\cdots [1]_q,\quad q=e^{\eta},\no \\[4pt]
&&B^{-}=\sum_{l=1}^Ne^{\frac{(N-1)\eta}{2}}\,e^{\frac{\eta}{2}\sum_{k=l+1}^N\sigma^z_k}\,\sigma^-_l \,
e^{-\frac{\eta}{2}\sum_{k=1}^{l-1}\sigma^z_k}.\label{B-operator-1}
\eea
Through the similar calculation, we obtain the homogeneous left reference state
\bea
\langle\Omega|=\sum_{l=0}^{\infty}\langle0|\frac{\lt(C^{+}\rt)^l}{[l]_q!}
=\sum_{l=0}^{N}\langle0|\frac{\lt(C^{+}\rt)^l}{[l]_q!},\label{reference-state-l}
\eea
where the explicit expression of the operator $C^{+}$ is
\bea
C^+=\sum_{l=1}^Ne^{\frac{(N-1)\eta}{2}}\,e^{-\frac{\eta}{2}\sum_{k=l+1}^N\sigma^z_k}\,\sigma^+_l \,
e^{\frac{\eta}{2}\sum_{k=1}^{l-1}\sigma^z_k}.
\eea

Put all the inhomogeneous parameters be zero in the BAEs (\ref{BAEs}) and solve them, we obtain the values of Bethe roots $\{u_k\}$ and $\{\lambda_k\}$, which are shown in Table 1.
\begin{table}[!ht]
\caption{ Bethe roots, where $\eta=1,N=3$. }\label{roots}
\centering
\begin{tabular}{|c|c|}
\hline 
 $\{u_k\}$ & $\{\lambda_k\}$ \\ \hline
 $-1.637416729786854 + 1.570796326794897i $ & $-1.431625849182040 - 0.000000000000000i$ \\ \hline
 $-0.500000000000000 + 1.570796326794896i $ & $-0.500000000000000 - 0.000000000000000i$ \\ \hline
 $0.637416729786854 + 1.570796326794897i $ & $0.431625849182040 + 0.000000000000000i $ \\ \hline
 \end{tabular}
\end{table}
\begin{table}[!ht]
\caption{ Numerical result of the scalar product $\langle\Phi\{\lambda_k\}|\Phi\{\lambda_k\}\rangle$, where $\eta=1,N=3$. } \label{scalar product-1}
\centering
\begin{tabular}{|c|c|}
\hline
 ${\rm definition}$ & $0.003625763123158$ \\ \hline
 $\varphi_1(0)$ & $0.667228749898571$ \\ \hline
 $|P^{hom}|$ & $0.058012209970527$ \\ \hline
 ${\rm analytical \ formula}$ & $0.003625763123158$  \\ \hline
\end{tabular}
\end{table}
\begin{table}[!ht]
\caption{ Numerical result of the form factor $\langle\Phi\{u_k\}| \sigma_{1}^{z} |\Phi\{\lambda_k\}\rangle$, where $\eta=1,N=3$. } \label{Fz}
\centering
\begin{tabular}{|c|c|}
\hline
 ${\rm definition}$ & $0.200953522733016i$ \\ \hline
 $\varphi_1(0)$ & $4.041937264439135i$ \\ \hline
 $\tilde{\xi}(0)$ & $0$ \\ \hline
 $\Lambda(\{u_k\},0)$ & $1.000000000000000$ \\ \hline
 $\Lambda(\{\lambda_k\},0)$  & $-1.000000000000000$ \\ \hline
 $|F^{hom,z}|$ & $-3.215256363728254i$ \\ \hline
 ${\rm analytical \ formula}$ & $0.200953522733016i$  \\ \hline
\end{tabular}
\end{table}
\begin{table}[!ht]
\caption{ Numerical result of the form factor $\langle\Phi\{u_k\}| \sigma_{1}^{-} |\Phi\{\lambda_k\}\rangle$, where $\eta=1,N=3$. } \label{F-}
\centering
\begin{tabular}{|c|c|}
\hline
 ${\rm definition}$ & $0.113108828168255i$ \\ \hline
 $f^-_1(0)$ & $4.674571757211132i$ \\ \hline
 $\Lambda(\{u_k\},0)$ & $1.000000000000000$ \\ \hline
 $\Lambda(\{\lambda_k\},0)$  & $-1.000000000000000$ \\ \hline
 $|F^{hom,D}|$ & $-1.809741250692130i$ \\ \hline
 ${\rm analytical \ formula}$ & $0.113108828168258i$  \\ \hline
\end{tabular}
\end{table}
\begin{table}[!ht]
\caption{ Numerical result of the correlation function $\langle\Phi\{\lambda_k\}|\sigma_{1}^{-}\sigma_{2}^{-} |\Phi\{\lambda_k\}\rangle$, where $\eta=1,N=3$. } \label{CF--}
\centering
\begin{tabular}{|c|c|}
\hline
 ${\rm definition}$ & $0.001006270991793$ \\ \hline
 $f^--_1(0)$ & $0.587693133253497$ \\ \hline
 $\Lambda(\{\lambda_k\},0)$  & $-1.000000000000000$ \\ \hline
 $|F^{hom,DD}|$ & $0.016100335868683$ \\ \hline
 ${\rm analytical \ formula}$ & $0.001006270991793$  \\ \hline
\end{tabular}
\end{table}

Now, we are ready to calculate all the scalar products. The scalar product of two Bethe states $\langle\Phi\{\lambda_k\}|\Phi\{\lambda_k\}\rangle$,
the form factors $\langle\Phi\{u_k\}| \sigma_{1}^{z} |\Phi\{\lambda_k\}\rangle$ and $\langle\Phi\{u_k\}| \sigma_{1}^{-} |\Phi\{\lambda_k\}\rangle$,
and the correlation function $\langle\Phi\{\lambda_k\}|\sigma_{1}^{-}\sigma_{2}^{-} |\Phi\{\lambda_k\}\rangle$ are shown in Tables 2-5, respectively.
In the these Tables, the first row is the results obtained from the definition, while the last row is the results obtained from the analytical formula. We see that
these two results agree with each other very well.

\section{Conclusions}

In this paper, based on the inhomogeneous $T-Q$ relation and Bethe states obtained in \cite{Cao13,Zha15}, we investigate the scalar products, the form factors and the two-point correlation functions of the XXZ spin chain with twisted boundary condition.
We find that the correlation function $\langle\Phi\{u_k\}|\sigma_{i-1}^-\sigma_i^-|\Phi\{\lambda_k\}\rangle$ can be expressed as a ratio of two determinants, while the correlation function
$\langle\Phi\{u_k\}|\sigma_{i-1}^z\sigma_i^z|\Phi\{\lambda_k\}\rangle$ can be expressed as a linear combination of $N+2$ determinants.
We also study the homogeneous limits of these results and check them numerically.
The analytical results and the numerical ones agree with each other very well. These results could be used to study the correlation length, critical behavior and dynamic properties of  the system.

\section*{Acknowledgments}
\setcounter{equation}{0}

The financial supports from the National Program
for Basic Research of MOST (Grant Nos. 2016YFA0300600 and
2016YFA0302104), the National Natural Science Foundation of China
(Grant Nos. 11434013, 11425522, 11547045, 11774397, 11775178 and 11775177), the Major Basic Research Program of Natural Science of Shaanxi Province
(Grant Nos. 2017KCT-12, 2017ZDJC-32), Australian Research Council (Grant No. DP 190101529) and  the Strategic Priority Research Program of the Chinese
Academy of Sciences, and the Double First-Class University Construction Project of Northwest University are gratefully acknowledged.



\begin{thebibliography}{99}

\bibitem{Kor93} V. E. Korepin, N. M. Bogoliubov and A. G. Izergin, Quantum Inverse Scattering Method and Correlation Function (Cambridge University Press, 1993).
\bibitem{3Baxter 2} R. J. Baxter, Exactly Solved Models in Statistical Mechanics (Academic Press, 1982).
\bibitem{Tak99} M. Takahashi, Thermodynamics of One-Dimensional Solvable Models (Cambridge University Press, 1999).

\bibitem{Yun95} C. M. Yung and M. T. Batchelor, Exact solution for the spin-$s$ XXZ quantum chain with non-diagonal twists, Nucl. Phys. B 446, 461 (1995).
\bibitem{Bat95} M. T. Batchelor, R. J. Baxter, M. J. O'Rourke and C. M. Yung, Exact solution and interfacial tension of the six-vertex model with anti-periodic boundary conditions, J. Phys. A. Math. Gen. 28, 2759 (1995).
\bibitem{Nie09} S. Niekamp, T. Wirth and H. Frahm, The XXZ model with anti-periodic twisted boundary conditions, J. Phys. A Math. Theor. 42, 195008 (2009).
\bibitem{Nic13} G. Niccoli, Antiperiodic spin-$1/2$ XXZ quantum chains by separation of variables: complete spectrum and form factors, Nucl. Phys. B 870, 397 (2013).



\bibitem{Che84} I. Cherednik, Factorizing particles on a half line and root sytsems, Theor. Math. Phys. 61, 977 (1984).
\bibitem{Skl88} E. K. Sklyanin, Boundary conditions for integrable quantum systems, J. Phys. A: Math. Gen. 21, 2375 (1988).
\bibitem{Fan96} H. Fan, B. -Y. Hou, K. Shi and Z. -X. Yang, Algebraic Bethe Ansatz for the eight-vertex model with general open boundary conditions, Nucl. Phys. B 478, 723 (1996).
\bibitem{Cao03} J. Cao, H. -Q. Lin, K. Shi and Y. Wang, Exact solution of XXZ spin chain with unparallel boundary fields, Nucl. Phys. B 663, 487 (2003).
\bibitem{Nep04} R. I. Nepomechie, Bethe ansatz solution of the open XXZ chain with nondiagonal boundary terms, J. Phys. A: Math. Gen. 37, 433 (2004).

\bibitem{10L.A. Takhtadzhan} L.A. Takhtadzhan and L.D. Faddeev, The quantum method of the inverse problem and the Heisenberg XYZ model, Rush. Math. Surv. 34, 11 (1979).
\bibitem{Skl78} E. K. Sklyanin and L. D. Faddeev, Quantum mechanical approach to completely integrable field theory models, Sov. Phys. Dokl. 23, 902 (1978).

\bibitem{Wan15} Y. Wang, W. -L. Yang, J. Cao and K. Shi, Off-Diagonal Bethe Ansatz for Exactly Solvable Models (Springer Press, 2015).



\bibitem{Ize87}A. G. Izergin, Partition function of the six-vertex model in a finite volume, Sov. Phys. Dokl. 32, 878 (1987).
\bibitem{Sla89}N. A. Slavnov, Calculation of scalar products of wave functions and form-factors in the framework of the algebraic Bethe ansatz, Theoret. and Math. Phys. 79, 502 (1989).
\bibitem{Gau81}M. Gaudin, B. M. McCoy and T. T. Wu, Normalization sum for the Bethe's hypothesis wave functions of the Heisenberg Ising chain, Phys. Rev. D 23, 417 (1981).
\bibitem{Kor82}V. E. Korepin, Calculation of norms of Bethe wave functions, Comm. Math. Phys. 86, 391 (1982).
\bibitem{Mai00}J. M. Maillet and V. Terras, On the quantum inverse scattering problem, Nucl. Phys. B 575, 627 (2000).
\bibitem{Goh00}F. G\"{o}hmann and V. E. Korepin, Solution of the quantum inverse problem, J. Phys. A 33, 1199 (2000).
\bibitem{Kit99}N. Kitanine, J. M. Maillet and V. Terras, Form factors of the XXZ Heisenberg spin-1/2 finite chain, Nucl. Phys. B 554, 647 (1999).
\bibitem{Kit09}N. Kitanine, K. K. Kozlowski, J. M. Maillet, N. A. Slavnov and V. Terras, On the thermodynamic limit of form factors in the massless XXZ Heisenberg chain, J. Math. Phys. 50, 095209 (2009).

\bibitem{Yan11}W. -L. Yang, X. Chen, J. Feng, K. Hao, B. Y. Hou, K. Shi and Y. -Z. Zhang, Determinant formula for the partition function of the six-vertex model with a non-diagonal reflecting end, Nucl. Phys. B 844, 289 (2011).
\bibitem{Yan11-1}W. -L. Yang, X. Chen, J. Feng, K. Hao, K. -J. Shia, C. -Y. Sun, Z. -Y. Yang and Y. -Z. Zhang, Domain wall partition function of the eight-vertex model with a non-diagonal reflecting end, Nucl. Phys. B 847, 367 (2011).
\bibitem{Kun12}K. Hao, W. -L. Yang, H. Fan, S. -Y. Liu, K. Wu, Z. -Y. Yang and Y. -Z. Zhang, Determinant representations for scalar products of the XXZ Gaudin model with general boundary terms. Nucl. Phys. B 862, 835 (2012).
\bibitem{Nep03}R. I. Nepomechie, Bethe ansatz solution of the open XXZ chain with nondiagonal boundary terms, J. Phys. A 37, 433 (2004).
\bibitem{Kit07}N. Kitanine, K. K. Kozlowski, J. M. Maillet, G. Niccoli, N. A. Slavnov and V. Terras, Correlation functions of the open XXZ chain I, J. Stat. Mech. P10009 (2007).
\bibitem{Fal14}S. Faldella, N. Kitanine and G. Niccoli, The complete spectrum and scalar products for the open spin-$1/2$ XXZ quantum chains with non-diagonal boundary terms, J. Stat. Mech. P01011 (2014).
\bibitem{Bel15}S. Belliard, R. Pimenta, Slavnov and Gaudin formulas for models without $U(1)$ symmetry: the twisted XXX chain, SIGMA 11, 099 (2015).



\bibitem{Cao13} J. Cao, W. -L. Yang, K. Shi and Y. Wang, Off-diagonal Bethe Ansatz and exact solution of a topological spin ring, Phys. Rev. Lett. 111, 137201 (2013).

\bibitem{Skl85} E. K. Sklyanin, The quantum Toda chain, Lect. Notes Phys. 226, 196 (1985).
\bibitem{Fra08} H. Frahm, A. Seel and T. Wirth, Separation of variables in the open XXX chain, Nucl. Phys. B 802, 351 (2008).


\bibitem{Qia18}Y. Qiao, Z.-R. Xin, K. Hao, J. Cao, W.-L. Yang, K. Shi and Y. Wang, Twisted boundary energy and low energy excitation of the XXZ spin torus at the ferromagnetic region, New J. Phys. 20, 073046 (2018).
\bibitem{Zha15} X. Zhang, Y.-Y. Li, J. Cao, W. -L. Yang, K. Shi and Y. Wang, Retrive the Bethe states of quantum integrable models solved via off-diagonal Bethe Ansatz, J. Stat. Mech. P05014 (2015).

\bibitem{Ize92} A. G. Izergin, D. A. Coker and V. E. Korepin, Determinant formula for the six-vertex model, J. Phys. A: Math. Gen. 25, 4315 (1992).
\end{thebibliography}
\end{document}